\documentstyle[aps,12pt]{revtex}

\begin{document}
\title{Non-Integrability and Chaos \\
in Classical Cosmology}
\author{Amina Helmi $ ^{(1)}$ $ ^{(2)}$\thanks{ahelmi@strw.LeidenUniv.nl}
and H\'{e}ctor Vucetich $ ^{(2)}$ \thanks{Member of the Carrera del Investigador
de CONICET.}} 

\address{$^{(1)}$ Sterrewacht Leiden, University of Leiden, Niels 
Bohrweg 2, 2333 CA Leiden, The Netherlands.}

\address{ $ ^{(2)}$ Facultad de Ciencias Astron\'omicas y
 Geof\'{\i}sicas, Universidad Nacional de La Plata, \\ Paseo del Bosque s/n,
1900 La Plata, Argentina.}

\maketitle

\begin{abstract}

A brief analysis of  the dynamics of a \mbox{Friedmann-Robertson-Walker} 
universe with a conformally coupled, real, \mbox{self-interacting}, 
massive scalar field, 
based  on the Painlev\'e theory of differential 
equations, is presented.
Our results complete earlier works
done within the framework of Dynamical System Theory.

We conclude that, in general, the system will not be integrable and that
the chaos that has been found in a previous work, arises from the presence of 
movable logarithmic branch 
points in the solution in the complex plane of time.

\vspace{0.2cm}
{\sl PACS:} 02.30.Hq, 04.40.Nr, 98.80.Cq
\end{abstract}

\vspace{1.2cm}

Several works in the latter years have tried to prove the importance
of non-linearities in General Relativity \cite{Rugh,Helmi,Calz1,Calz2}
that may lead to the onset of chaos in the early stages of the
universe. The chaotic behavior is even observed to occur in quite
simplistic models like vacuum Bianchi IX (see
\cite{Barrow,Contopoulos,Latifi}), due to only the coupling of the
different scale factors, as this model does not contain any matter
fields.  The focus has also been put on the inflationary phase of the
early universe, and on the role that different fields might have
played in this scenario \cite{Linde1,Kung,Calz1}.  In many of these
papers, the problem was attacked by means of numerical tools, based on
Dynamical System Theory, which as we will show later in this brief
report, do not explore some of the very interesting features of the
system under study.  Moreover, some people have shown (see \cite{Rugh}
for a discussion of the problem) that the detection of chaotic
behavior is quite dependent on the choice of the coordinate system,
and thus methods which provide gauge independent (or covariant)
measures or results about the behavior of the system under study are
very much needed while working within theories of Gravitation.  Lately
\cite{Levin} studies based on fractal basin boundaries have shown to
be very useful, as no smooth coordinate transformation can undo a
fractal pattern.  On the other hand, from the analytical point of
view, singularity studies can provide with very meaningful results,
which are independent of any particular gauge choice.  And even more,
they can be shown to be invariant under the group of homographic
transformations of the dependent variables \cite{Conte}.

In this report we study the dynamics of a conformally coupled field in
a homogeneous and isotropic universe. We shall consider a massive,
self-interacting, scalar field conformally coupled to a
Friedmann-Robertson-Walker metric \cite{Kolb}.  The system is
described by the following two degrees of freedom Hamiltonian:

\begin{equation}
\label{Hamilt}
H = \frac{1}{2} [-(p_{a}^{2} + k a^{2}) + (p_{\phi}^{2} + k \phi^{2}) + 
m^{2}a^{2}\phi^{2} + \frac{\lambda}{2} \phi^{4} + \frac{\Lambda}{2} a^{4} ]
\end{equation}
 
where $a$ is the scale factor, $k$ is the curvature and $\Lambda =
\frac{2 \Lambda_{0}}{3}$ is related to the cosmological constant. We
have obviously considered a potential \mbox{$V(\Phi)=
\frac{1}{2}m^{2}\Phi^{2} + \frac{1}{4}\hat{\lambda}\Phi^{4}$,} where
\mbox{$\Phi = \phi/(a \sqrt{v})$,} \mbox{$\lambda = \hat{\lambda}/v$}
is the \mbox{self-interacting} coefficient ($v$ is the conformal
volume) and $m$ is the mass of the field.  \mbox{$m^{2}$ can} take
both positive and negative values, what would take into account
processes of spontaneous symmetry breaking resulting in phase
transitions which may have occurred in the early stages of the
\mbox{universe \cite{Guth,Linde2,Albrecht,Kolb2}.}

The Hamiltonian presented above (eq.~(\ref{Hamilt})) is well-known and
has been thoroughly studied within lattice-dynamics, condensed matter
theory, field theory, etc. (see references \mbox{in \cite{Lak}).} It
is interesting to note that if we make the change of variable $ a = -i
x$, the Hamiltonian would correspond to that of a set of two
quartically coupled non-linear \mbox{oscillators \cite{Lak},} the
study of which can be done analytically, giving a more complete view
of the behavior of the system.

The analytical method is based on the Painlev\'e analysis of the
differential equations, which states that if a solution is free from
movable critical points other than poles, then it can be expressed as
a generic Laurent series around one of its singularities. The test
known as ARS algorithm \cite{Ablowitz}, goes through the analysis of
the leading order terms, the next order terms (or resonances) and the
number of arbitrary constants that can be introduced in the expansion
so as to make it generic. (It is important to note that the test gives
only necessary conditions for the integrability of the system.)

When one studies the dominant behaviors of the system derived from eq.~(1)
\begin{eqnarray}
\ddot{a} + k a - m^2 a \phi^2 - \Lambda a^3 & = & 0, \\
\ddot{\phi} + k \phi + m^2 a^2 \phi + \lambda \phi^3 & = & 0,
\end{eqnarray}
it is
immediately possible to see that the leading order terms are
independent of the value of the curvature $k$ (see \mbox{TABLE
\ref{tab:1}),} which is a very interesting and important result, as we
shall interpret below. 
The cases for which the system passes P-test are very few. They correspond to
special values of the parameters such as
\begin{eqnarray}
  & \left\{\begin{array}{l} 
 \Lambda = \lambda \;\; \mbox{and} \;\; m^2 = - 3 \Lambda, \\
 \Lambda = \lambda \;\; \mbox {and}\;\; m^2 = - \Lambda, \\ 
 \end{array}
 \right. \;\; \; & \forall k, \\
& \left\{\begin{array}{l}
\Lambda = 16 \lambda \;\;\mbox{and}\;\; m^2 = - 6 \lambda,  \\
\Lambda = 8 \lambda \;\;\mbox{and}\;\; m^2 = - 3 \lambda, 
\end{array}\right.\;\;\; & \mbox{for $k = 0$.}
\end{eqnarray}
In \mbox{reference \cite{Lak}} the integrals of motion are given
for these cases.

For the rest of the cases, which are the majority , the system shows
different types of movable critical points: algebraic branch points,
with irrational or complex values for the exponents, or logarithmic
branch points. (See TABLE \ref{tab:2}. Notice that, indeed, for a very
few values of the parameters {\sf Branch 1} will have integer
resonances.  In general the resonances will be irrational or complex,
thus leading to the nonintegrability of the system. The same would
happen in {\sf Branch 2}, if ($r^{2}=1 - 8m^{2}/\lambda < 0$), or if
$r^{2}$ is not the square of an integer or rational number. The same
of course applies to {\sf Branch 3}.)

It is interesting to compare these results with those obtained by
Blanco et al. \cite{chaos1,chaos2} by means of numerical methods. They
analyzed the same Hamiltonian, but restricted themselves to the cases
for which $k \neq 0$ (recall that $k = 0$ is a very important case
indeed) and to a particular choice for the values of the parameters ($
2m^{2} + \lambda + \Lambda= 0$).  The first restriction comes from a
stability analysis which shows that the only fixed point for $k = 0$
is an elliptic point. This lead them to conclude, erroneously as we
shall show, that a chaotic behavior cannot be reached in this flat
universe.  The second condition is a consequence of the Hamiltonian
constraint $H = 0$ in General Relativity. With our approach, this
constraint is always satisfied for each of the leading orders
considered, without having to impose any restrictions on the values of
the parameters.  It is very easy to show with the method that we have
been using, that the system passes the first two steps of the
algorithm (leading order corresponds to {\sf Branch 1}, the resonances
are $-1$ and $4$ with double multiplicity), but that the intented
Laurent expansion will not have a sufficient number of arbitrary
constants unless one introduces logarithmic terms in it. This fact is
seen in the following: $-a_{0}^{2} = b_{0}^{2} = \frac{4}{|\Lambda -
\lambda|}$, $ia_{2}= \frac{k (\Lambda^{2} + \lambda^{2} - 6\Lambda
\lambda)}{6 (5 \Lambda^{2} + \lambda^{2} - 6 \Lambda \lambda)}
(-a_{0}), \, b_{2} = \frac{kb_{0}}{6}$, $a_{1}=a_{3}=b_{1}=b_{3}=0$,
$a_{4}$ is arbitrary, and $b_{4}= f(a_{4})$. Thus a general solution
would be of the form: $ x(t)= ia(t) =a_{0} \tau^{-1} + a_{2} \tau +
(a_{4} + c_{4} \log{\tau})\tau^{4} + ... $, and a similar expansion
for $\phi(t)$.

Moreover, the chaos that has been observed by Blanco et al., can
immediately be related to the presence of these movable logarithmic
branch points.  It is well known \cite{Ramani,Bountis}, that systems
possessing this type of critical points display very interesting
features in the complex plane of time. The singularities tend to
accumulate in the complex $t-$plane, to form characteristic patterns
such as chimneys (or spirals with increasing and decreasing radii for
the case of complex resonances). It is also possible to show that the
distance between neighboring singularities steadily decreases. Because
of this, it is impossible to extend the solution further than a
certain distance from a pair of initial singularities. Any path of
analytic continuation appears to be trapped in a geometrically
converging web of singularities that creates a natural boundary of the
solution \cite{Chang1,Chang2}.

In synthesis, the above study has shown that there exists a relation
between the occurrence of movable branch points (i.e. multivaluedness,
formation of self-similar patterns which eventually will be natural
boundaries if they lie on the same Riemann sheet) and the onset of
chaos in a system.  We should emphasize again that this result is
independent of the value of $k$, and includes the case of a flat
universe. This shows that there is no reason to believe that chaos
will not be reached in this case.

\vspace{0.5cm}
 
Our results, based on the Painlev\'e analysis, show that for most of
the values the parameters can take, the system is not integrable
because of the presence of movable branch points, either algebraic
(irrational and complex values for the resonances) or logarithmic.
Besides, our results are independent of the curvature, as might have
been expected, since it is well-known that the role of the curvature
term can be neglected in Friedmann's equation during the inflationary
phase.  Thus, chaos is likely to be developed for all 
initial conditions of an
homogeneous and isotropic universe, independently of the value of the
curvature, cosmological and coupling constants, except for a zero
measure set of values. The cases for which a chaotic stage may have 
never been reached are only a very few \cite{Gramm}, that include 
for example when
the coupling constant $\lambda$ is equal to the cosmological 
constant $\Lambda$.  This
fact would in some sense rule out the very well-known initial-value
problem of the very early states of the universe, if the time scale
for which the chaotic behavior is completed is comparable to the
characteristic time for which the inflationary phase is expected to
last. If during the final stages of $\Lambda$-dominated inflation the 
scalar field were to oscillate around the potential minimum in a chaotic way,
the particles produced in this process could be responsible for the 
subsequent reheating of the universe \cite{Linde3}. This reheating 
in a chaotic fashion 
might be then, the necessary one to produce a sufficiently high 
temperature so as to restore the scenario for the standard baryogenesis
to take place.

\acknowledgements

This work was partially supported by CONICET, UNLP and the Vatican
Observatory.  A. Helmi wishes to thank Zonta International for the
Amelia Earhart Fellowship Award.

\newpage

\begin{table}
\caption{Leading order terms: $x\approx a_{0}\tau^{p}$, 
$\phi \approx b_{0} \tau^{q}$ (with $\tau = t- t_{0}$)}
\label{tab:1}
\begin{tabular}{ccccc}
& $p$ & $q$ & $a_{0}^{2}$ & $b_{0}^{2}$ \\
\tableline
{\sf Branch 1} & $-1$ & $-1$ & $2 \frac{(m^{2} +
\lambda)}{m^{4}-\lambda\Lambda}$ & $2
\frac{m^{2}+\Lambda}{m^{4}-\lambda\Lambda}$  \\
{\sf Branch 2} & $\frac{1 \pm \sqrt{1 - 8 m^{2}/\lambda}}{2}$ & $-1$ & arbitrary & 
$ - 2/\Lambda$ \\
{\sf Branch 3} & $-1$ & $\frac{1 \pm \sqrt{1 - 8 m^{2}/\Lambda}}{2}$ & $-
2/\lambda$ & arbitrary \\
\end{tabular}
\end{table}

{\sf Branch 2} only appears if ($m^{2} > 0$ and $\lambda > 0$),  or if 
($ 0 < -m^{2} < \lambda$).
And {\sf Branch 3}, if ($m^{2} > 0$ and $\Lambda > 0$), or if 
($ 0 < -m^{2} < \Lambda$).

\begin{table}
\caption{Resonances: $x(t)=a_{0}\tau^{p} + a_{1}\tau^{p + r}$, etc.}
\label{tab:2}
\begin{tabular}{cc}
&  $r$ \\
\tableline
{\sf Branch 1} & $x = \frac{3}{2} \pm \frac{1}{2(m^{4} - \lambda \Lambda)} 
\sqrt{ - 7m^{8} - 16(\lambda + \Lambda) m^{6} - 18\lambda \Lambda m^{4} + 
16\lambda \Lambda (\lambda + \Lambda) m^{2} + 25 \lambda^{2} \Lambda^{2}}$,
$-1$,  $4$. \\
{\sf Branch 2} & $  \mp \sqrt{1  - 8m^{2}/\lambda},$ \,$-1$, \,$0$, \,$4$ \\
{\sf Branch 3} & $  \mp \sqrt{1  - 8m^{2}/\Lambda},$ \,$-1$, \,$0$, \,$4$ \\
\end{tabular}
\end{table}

\end{document}